\documentclass[
]{ceurart}

\graphicspath{{fig/}}

\begin{document}

\copyrightyear{2021}
\copyrightclause{Copyright for this paper by its authors.\\
  Use permitted under Creative Commons License Attribution 4.0
  International (CC BY 4.0).}

\conference{}
\title{News Information Decoupling: An Information Signature of Catastrophes in Legacy News Media}

\author[1,2,3]{Kristoffer L. Nielbo}[%
orcid=0000-0002-5116-5070,
]
\ead{kln@cas.au.dk}
\ead[url]{https://knielbo.github.io/}

\author[3]{Rebekah B. Baglini}[
orcid=0000-0002-2836-5867,
]
\ead{rbkh@cc.au.dk}

\author[1,2]{Peter B. Vahlstrup}[
orcid=0000-0003-4126-7740,
]
\ead{imvpbv@cc.au.dk}

\author[1]{Kenneth C. Enevoldsen}[
orcid=0000-0001-8733-0966,
]
\ead{kenneth.enevoldsen@cas.au.dk}

\author[2]{Anja Bechmann}[%
orcid=0000-0002-5588-5155,
]
\ead{anjabechmann@cc.au.dk}

\author[3]{Andreas Roepstorff}[
orcid=0000-0002-3665-1708,
]
\ead{andreas.roepstorff@cas.au.dk}

\address[1]{Center for Humanities Computing Aarhus, Jens Chr. Skous Vej 4, Building 1483, 3rd floor, DK-8000 Aarhus C, Denmark}

\address[2]{DATALAB, School of Communication and Culture, Aarhus University, Helsingforsgade 14, DK-8200 Aarhus N, Denmark}

\address[3]{Interacting Minds Centre, Jens Chr. Skous Vej 4, Building 1483, 3rd floor, DK-8000 Aarhus C, Denmark}

\begin{abstract}
Content alignment in news media was an observable information effect of Covid-19's initial phase. During the first half of 2020, legacy news media became `corona news' following national outbreak and crises management patterns. While news media are neither unbiased nor infallible as sources of events, they do provide a window into socio-cultural responses to events. In this paper, we use legacy print media from Denmark to empirically derive the  principle News Information Decoupling (NID) that functions as an information signature of culturally significant catastrophic event. Formally, NID provides input to change detection algorithms and points to several unsolved research problems in the intersection of information theory and media studies.  
\end{abstract}


\maketitle

\noindent ``Nothing travels faster than the speed of light with the possible exception of bad news, which obeys its own special laws.''
\\[5pt]
\rightline{{ --- Douglas Adams}}

\section*{Introduction}

As the first wave of Covid-19 virus spread across the world, content alignment of news stories could be observed both within and between media sources. During December 2019 and February 2020, Covid-19 news stories were, outside China, interspersed with news coverage of other events (e.g., Hong Kong protests, Iranian–American confrontation, Trump impeachment). As the virus spread across Europe and America, news media front pages focused almost exclusively on the pandemic, all news sections (politics, business, sports, and arts) related to Covid-19, and breaking news became \emph{corona news} in continuously updated media. From the perspective of cultural dynamics, the Covid-19 pandemic provides a natural experiment that allows us to study the effect of a global catastrophe on the the dynamics of news media's information. While news media are neither unbiased nor infallible as sources of events, they do reflect preferences, values, and desires of a wide socio-cultural and political user spectrum. As such, news media coverage of Covid-19 functions as a proxy for how cultural information systems respond to unexpected and dangerous events.

Previous studies have shown that variation in newspapers' word usage is sensitive to the dynamics of socio-cultural events \cite{guldi_measures_2019, van_eijnatten_eurocentric_2019, daems_workers_2019}, can detect event-driven shifts \cite{kestemont_mining_2014}, and accurately can model effects of change in comprehensive collections of newspapers \cite{bos_quantifying_2016}. Furthermore, the co-occurrence structure of newspapers has been shown to accurately capture thematic development \cite{newman_probabilistic_2006}, and, when modelled dynamically, is indicative of the evolution of cultural values and biases \cite{van_eijnatten_eurocentric_2019, wevers_using_2019}. Adaptive smoothing and fractal analysis of word frequencies over time have been used to identify distinct domains of newspaper content (e.g., advertisements vs. articles) \cite{wevers_tracking_2020} and to discriminate between different classes of catastrophic events that display class-specific fractal signatures in, among other things, word usage in newspapers \cite{gao_culturomics_2012}. Several studies have shown that measures of (relative) entropy can detect fundamental conceptual differences between distinct periods \cite{guldi_measures_2019},  concurrent normative and ideological movements \cite{barron_individuals_2018}, and even, development of ideational factors (e.g., creative expression) in temporally dependent writings \cite{murdock_exploration_2015, nielbo_automated_2019, nielbo_curious_2019}. More specifically, a set of methodologically related studies studies have applied windowed relative entropy to thematic text representations to generate signals that capture information \emph{novelty} as a reliable content difference from the past and \emph{resonance} as the degree to which future information conforms to said novelty \cite{barron_individuals_2018, murdock_exploration_2015}. Two recent studies have found that successful social media content show a strong association between novelty and resonance \cite{nielbo_trend_2021}, and that variation in the novelty-resonance association can predict significant change points in historical data \cite{vrangbaek_composition_2021}.

In this study we expand upon studies of novelty and resonance in cultural dynamics by modeling change in printed news media during the initial phase of Covid-19. Specifically, we propose the empirically derived principle of News Information Decoupling, which explains how the information flow in legacy media responds to catastrophic events.

\begin{figure}
    \centering
    \includegraphics[width=.55\textwidth]{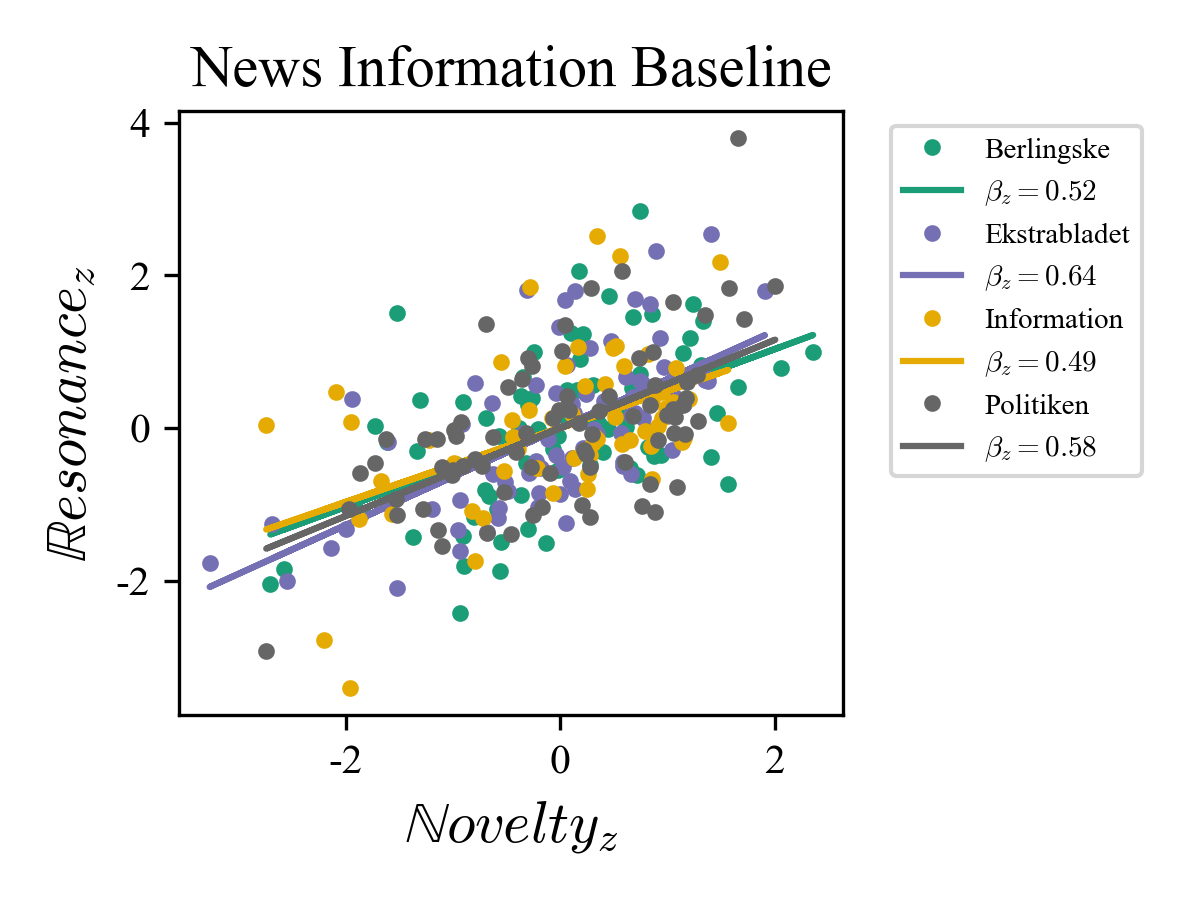}
    \caption{$\mathbb{N} \times \mathbb{R}$ slope baseline for four national newspapers that represents the left-right political spectrum. Data are sampled before Covid-19 phase 1 in Denmark initiated.}
    \label{fig:baseline}
\end{figure}

\begin{figure}
	\centering
	    \includegraphics[width=.75\textwidth]{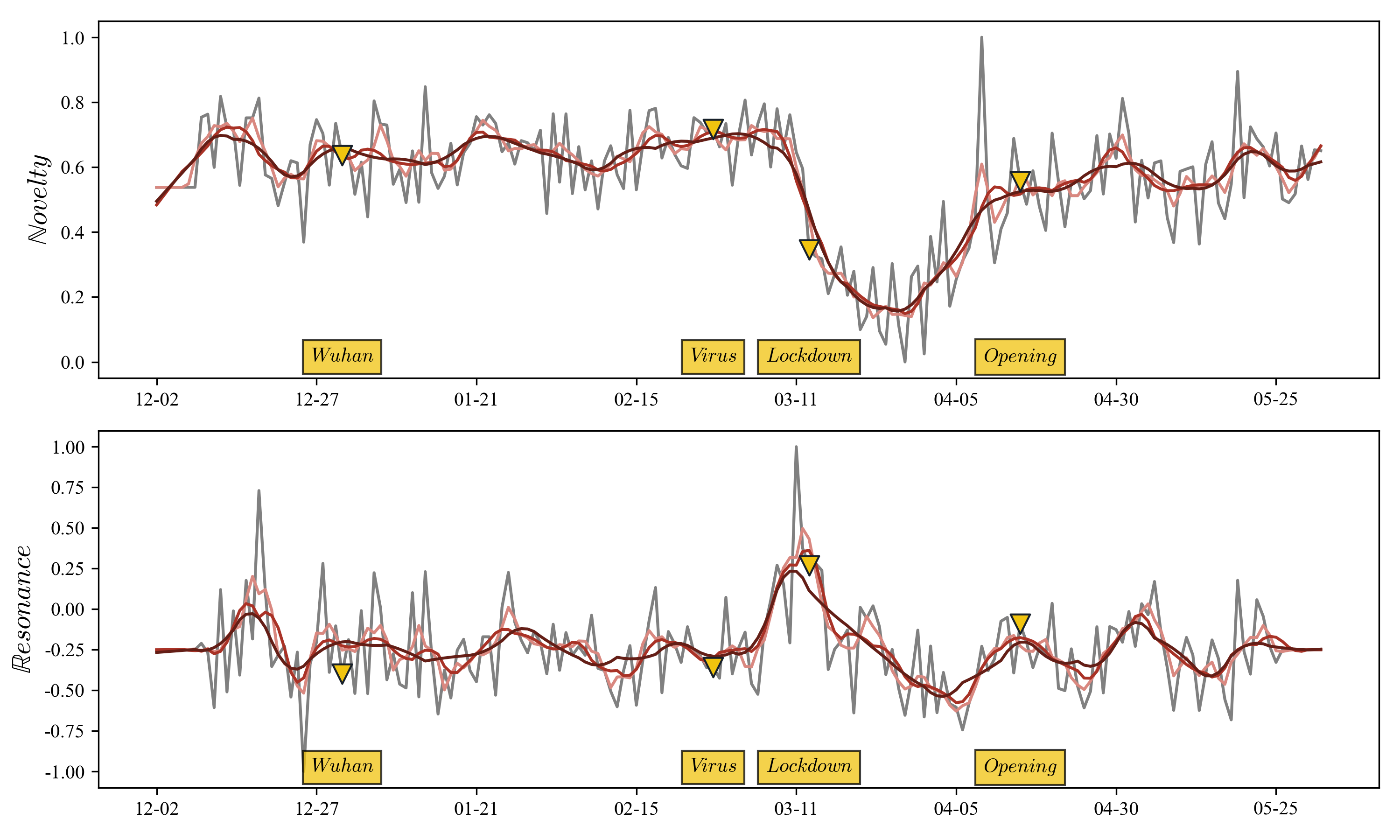}\\
	    \includegraphics[width=.19\textwidth]{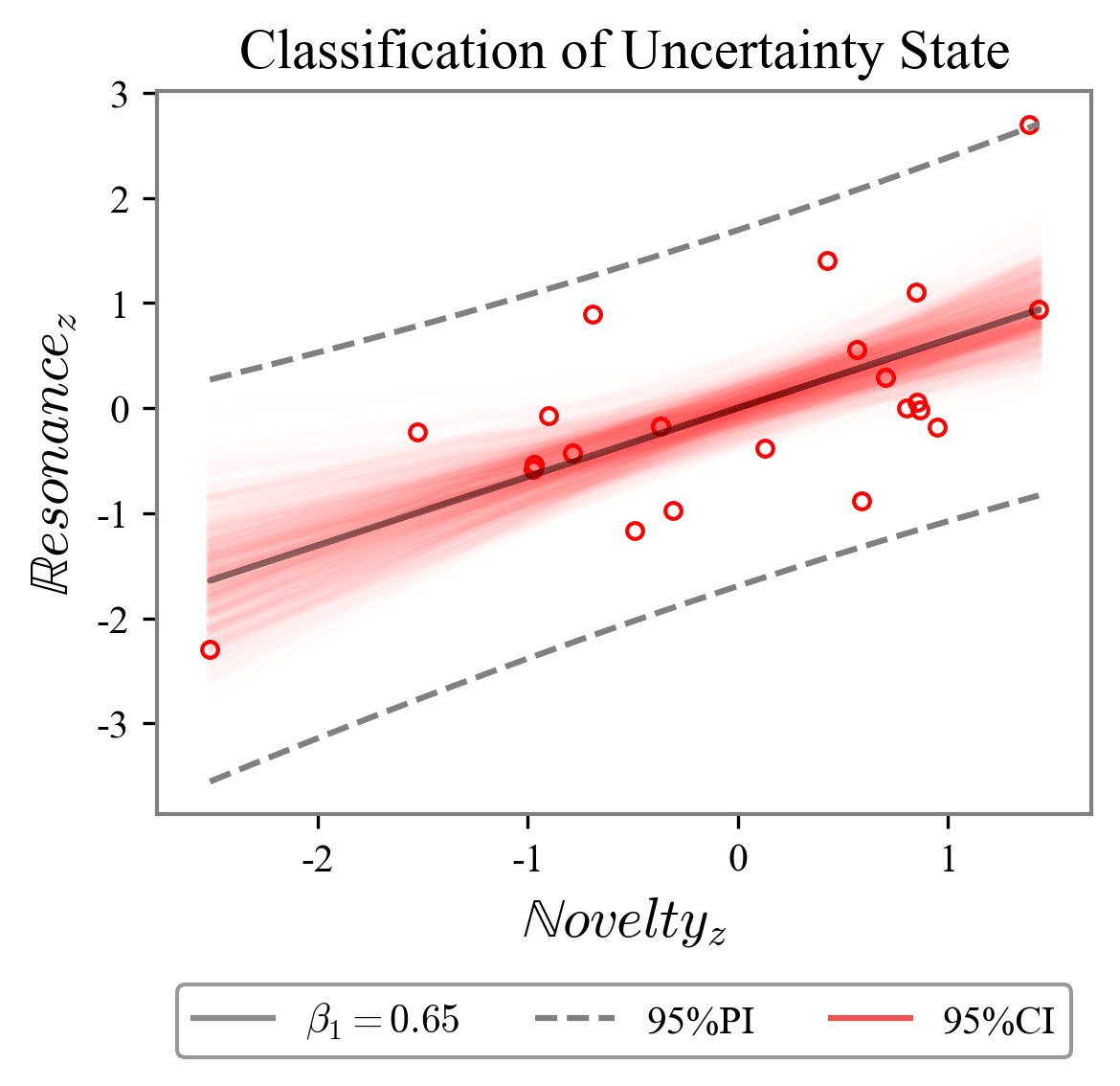}
		\includegraphics[width=.19\textwidth]{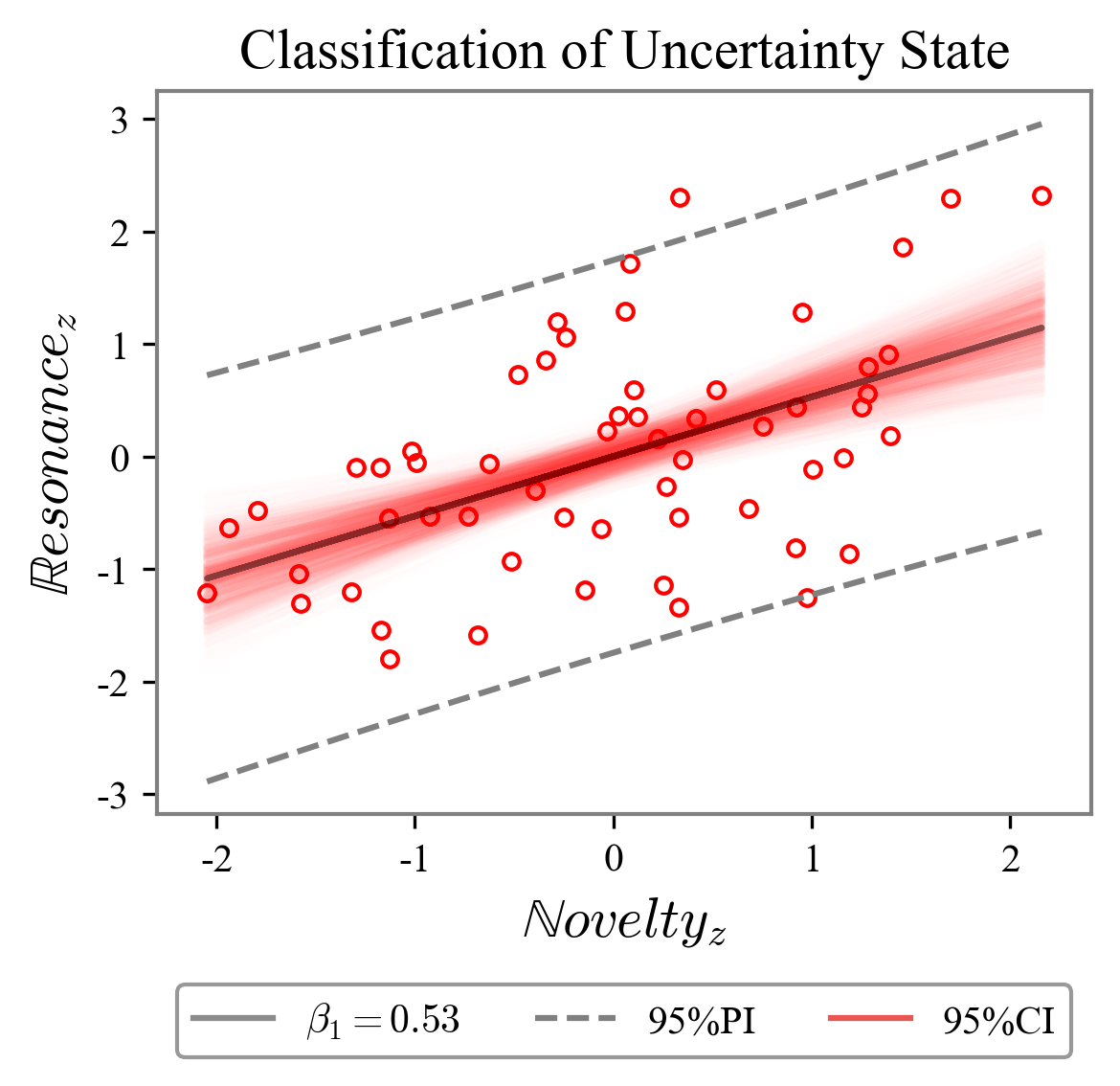}
		\includegraphics[width=.19\textwidth]{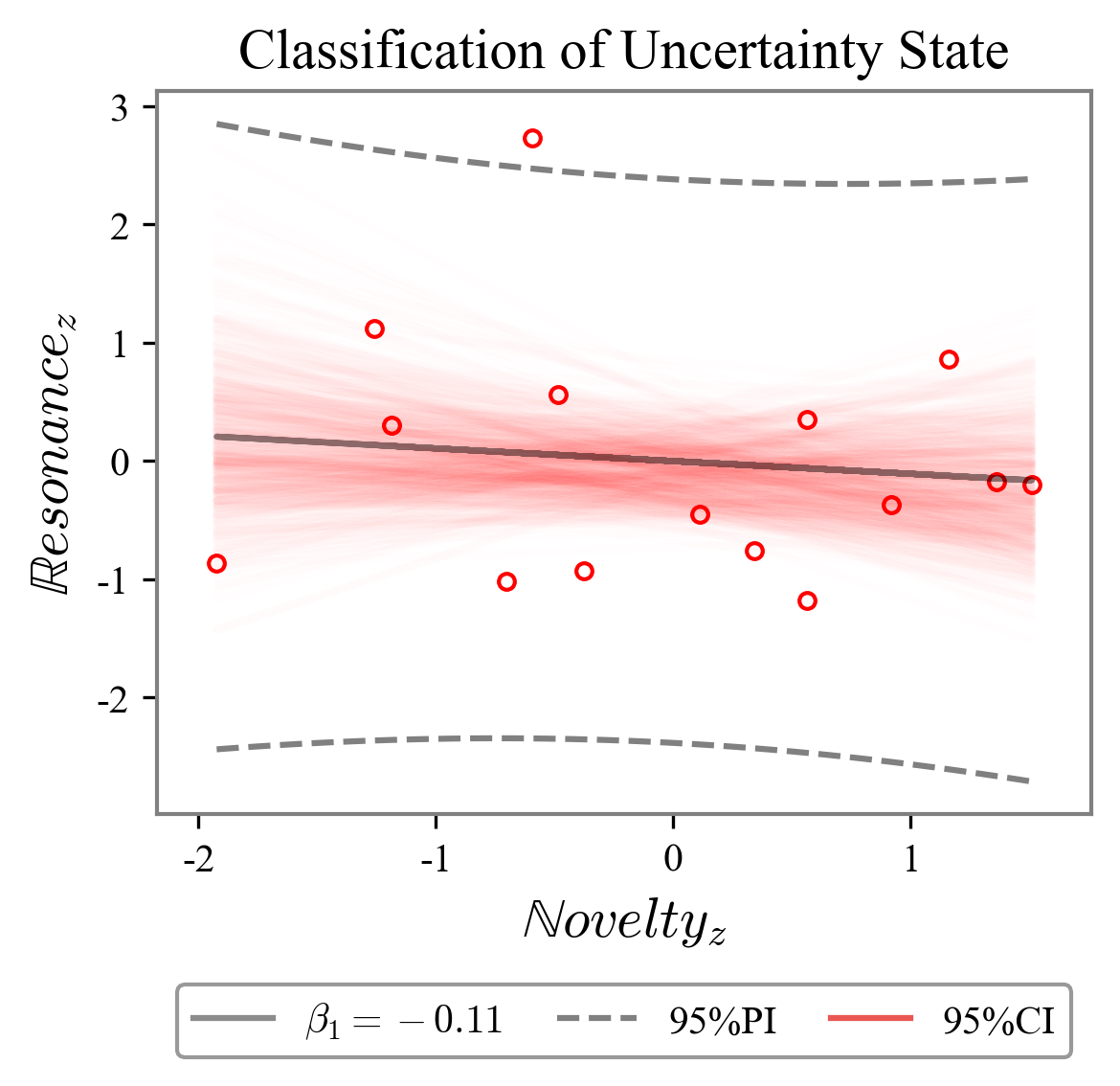}
		\includegraphics[width=.19\textwidth]{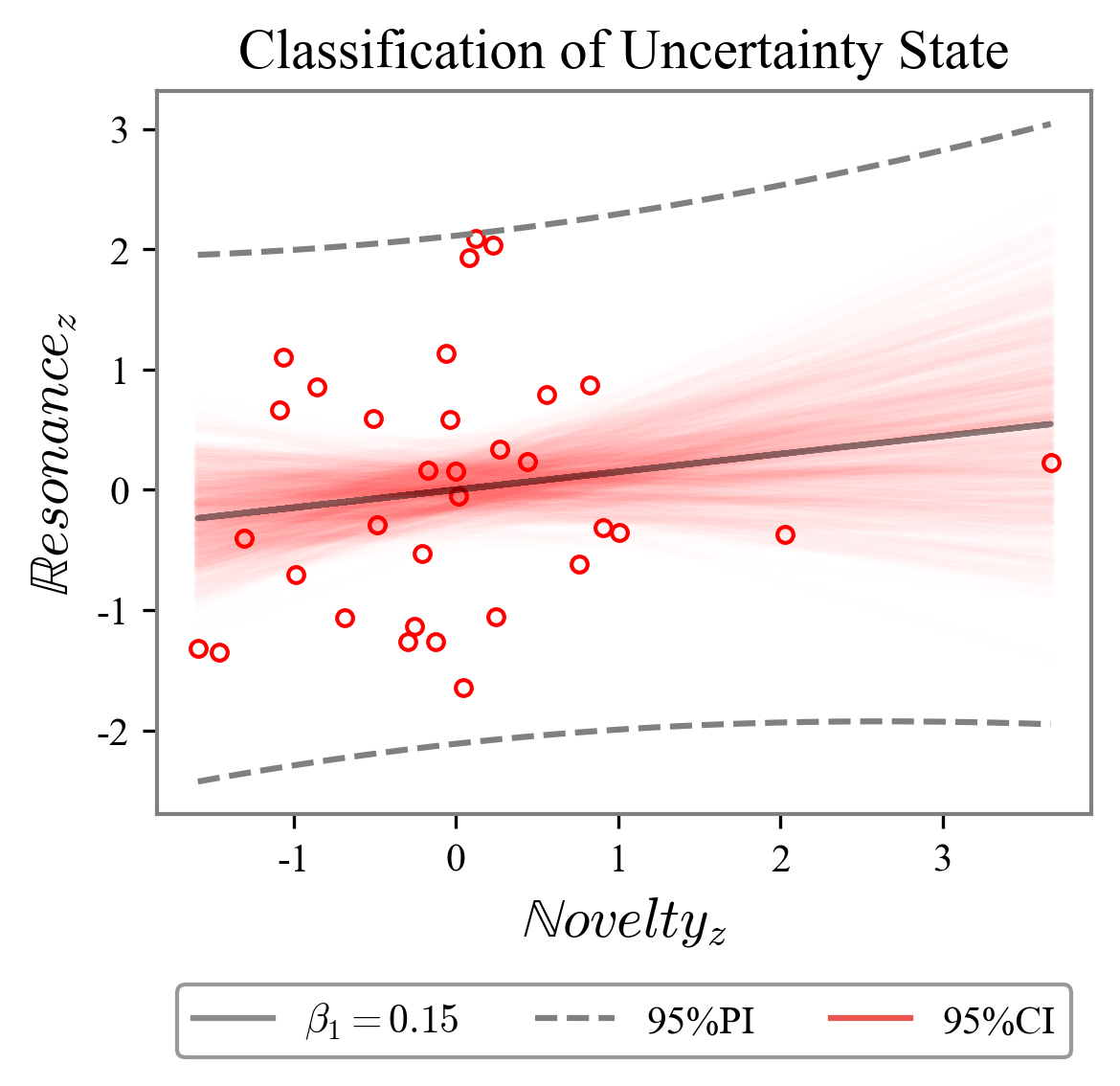}
		\includegraphics[width=.19\textwidth]{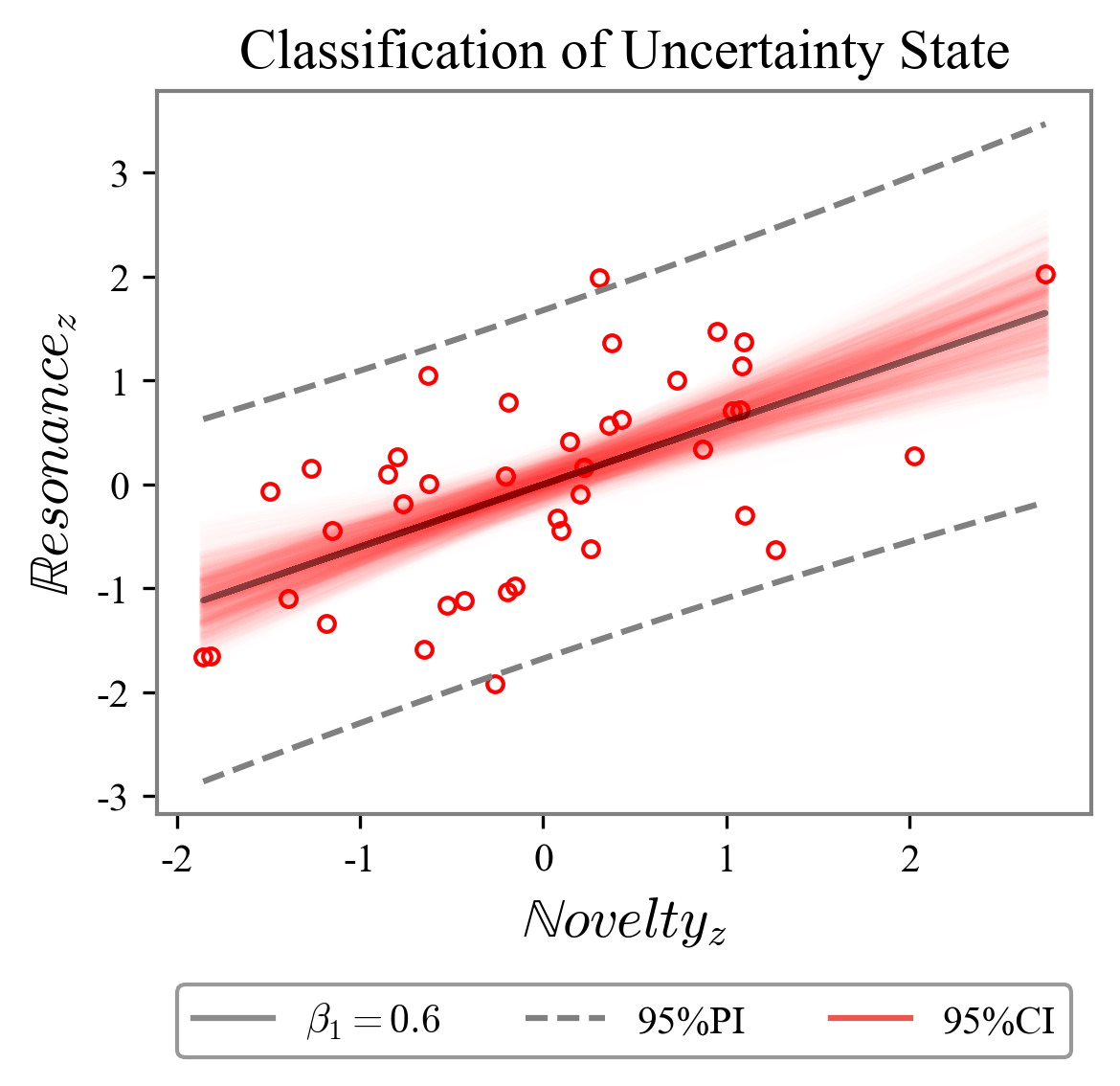}
	\caption{Novelty (upper panel), resonance (middle panel), and  event-defined $\mathbb{N} \times \mathbb{R}$ slopes (lower panel) for center-left newspaper \textit{Politiken} before and during Covid-19 phase 1. Trend lines in the upper and middle panel are estimated using a nonlinear adaptive filter, see appendix \ref{appA}.}
    \label{fig:poladapt}
\end{figure}

\begin{figure}
	\centering
	    \includegraphics[width=.75\textwidth]{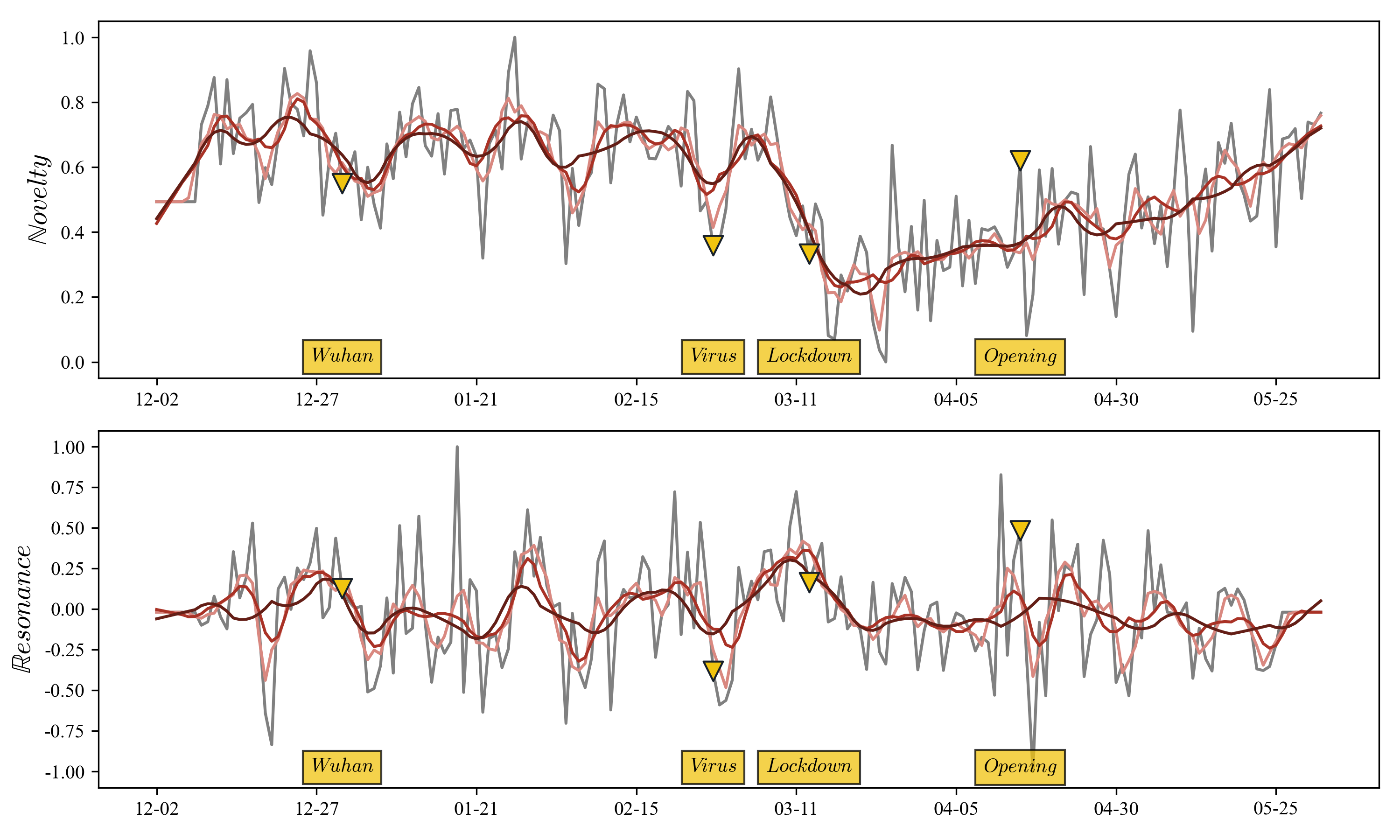}\\
	    \includegraphics[width=.19\textwidth]{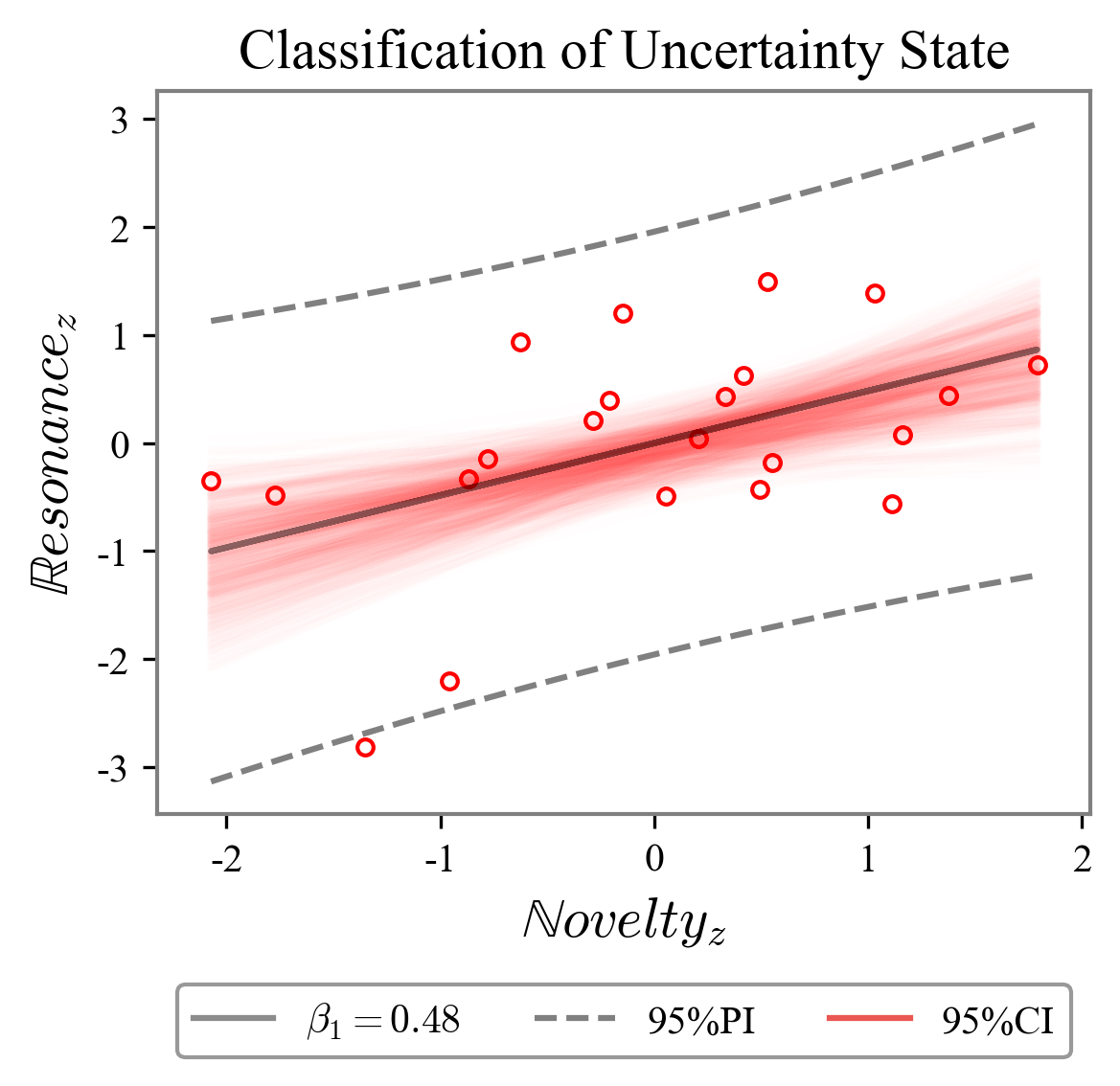}
		\includegraphics[width=.19\textwidth]{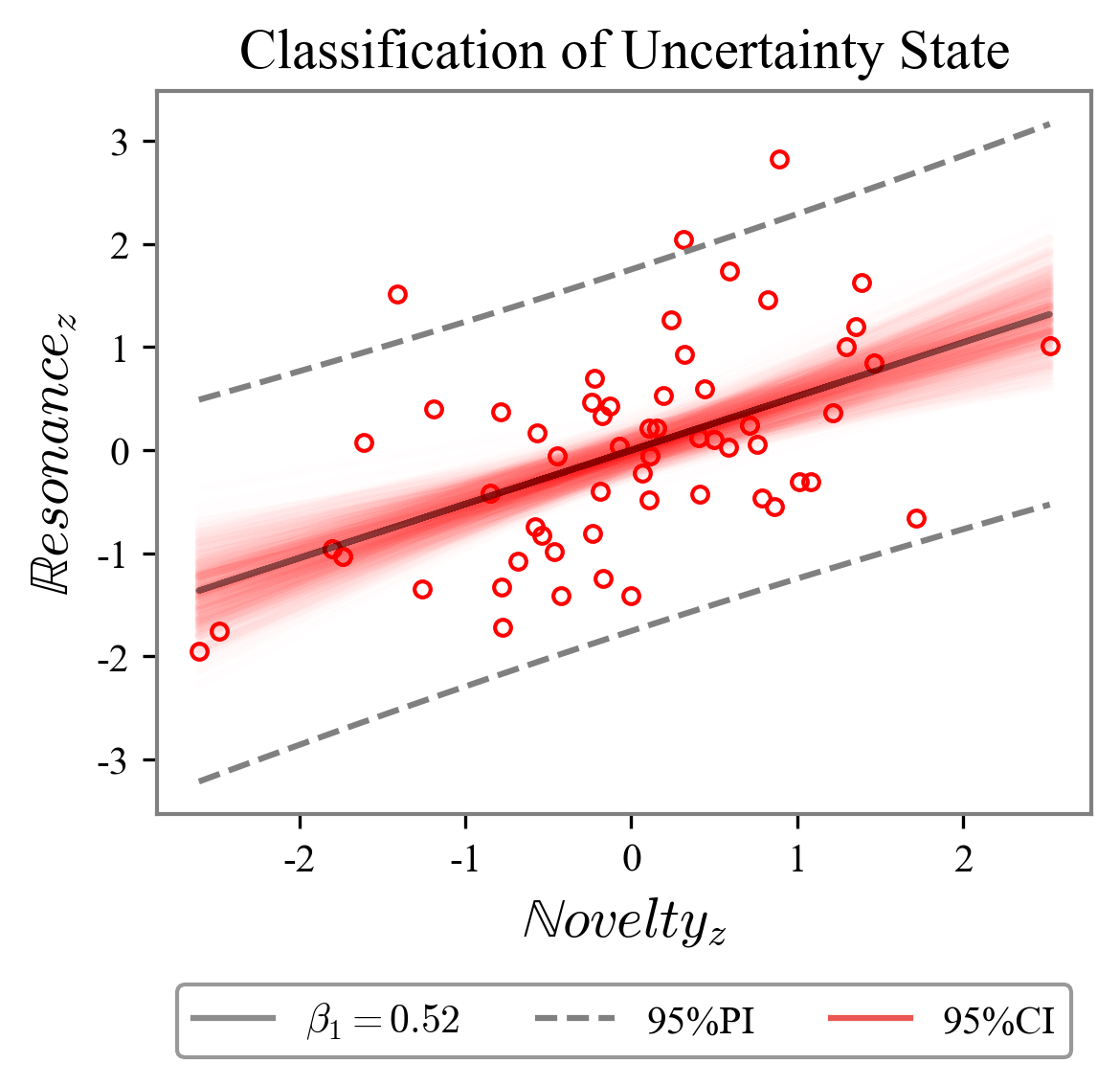}
		\includegraphics[width=.19\textwidth]{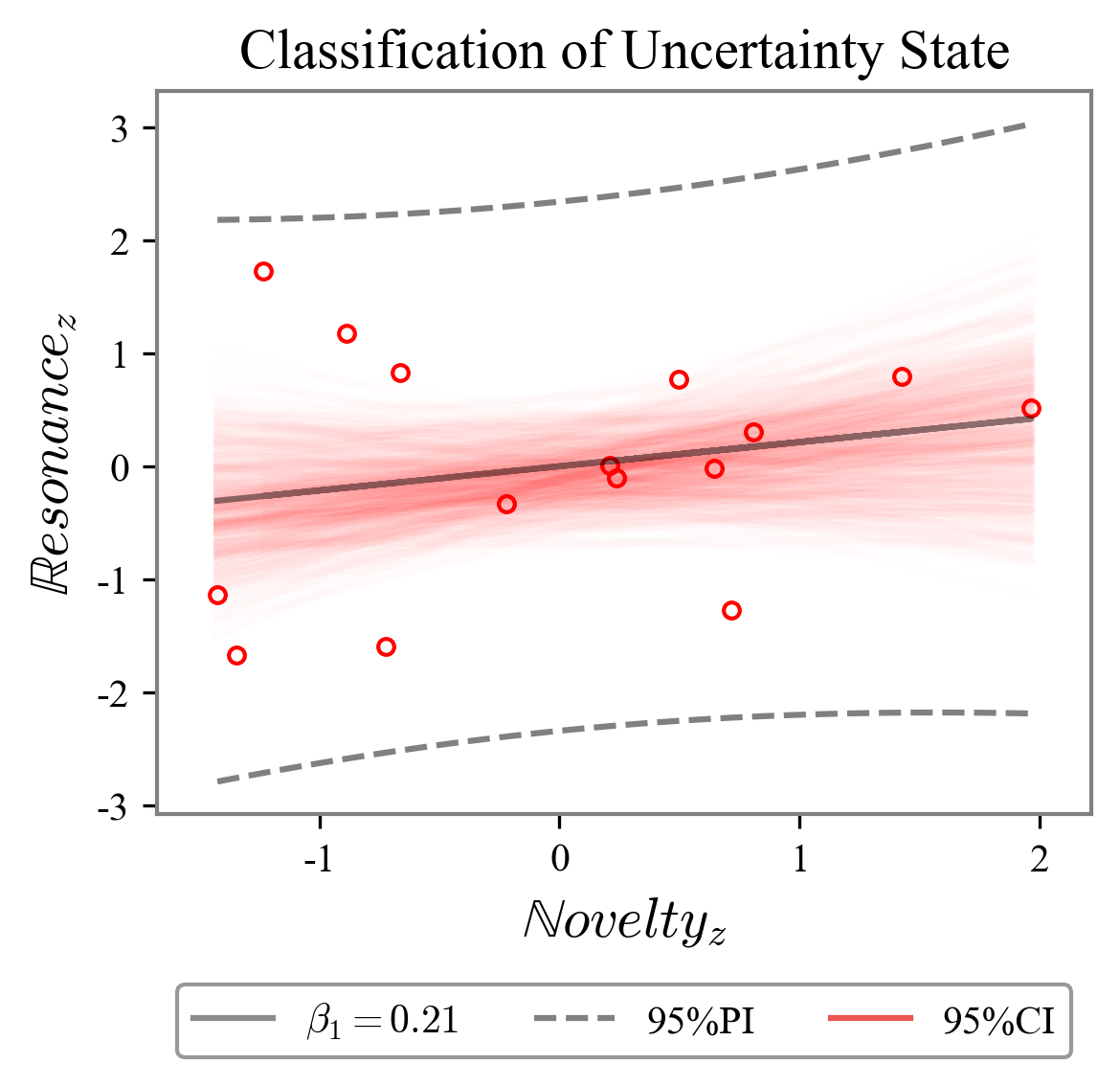}
		\includegraphics[width=.19\textwidth]{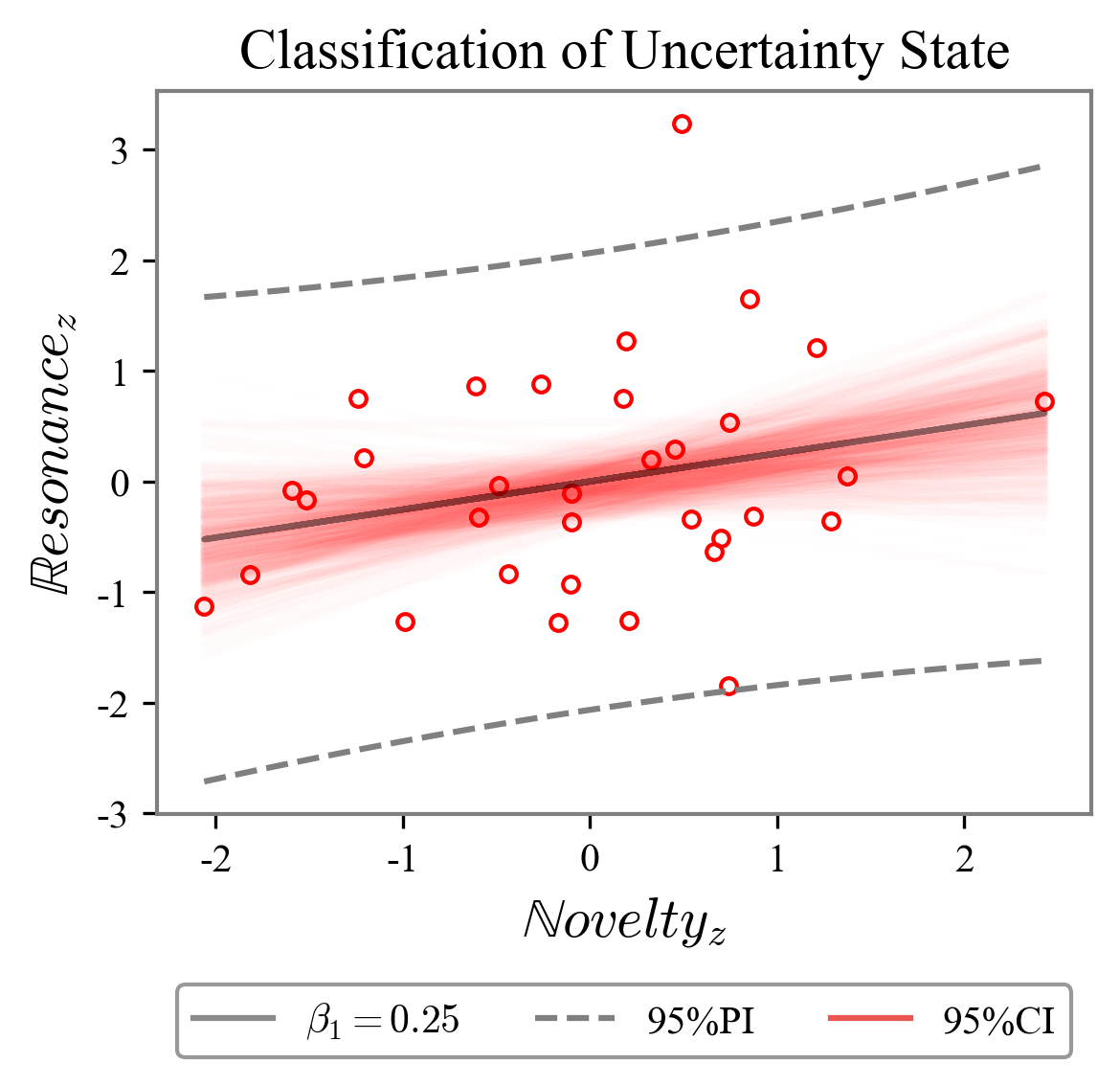}
		\includegraphics[width=.19\textwidth]{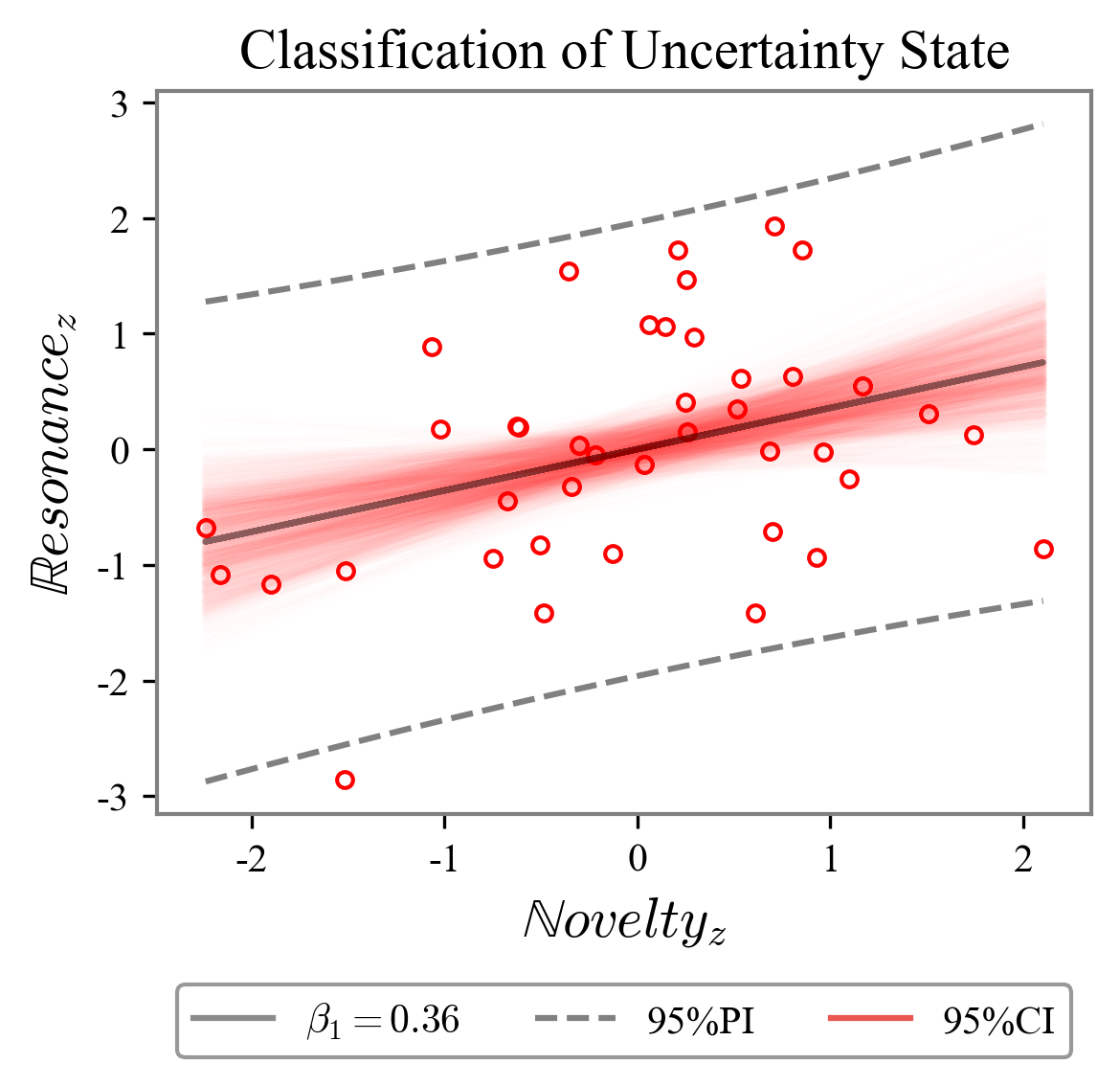}
	\caption{Novelty (upper panel), resonance (middle panel), and  event-defined $\mathbb{N} \times \mathbb{R}$ slopes (lower panel) for center-right newspaper \textit{Berlingske} before and during Covid-19 phase 1.}
    \label{fig:beradapt}
\end{figure}

\section*{Results}
The reported results are based on a sample of fours national newspapers in Denmark that collectively represents moderate-left (\textit{Politiken}) and moderate-right (\textit{Berlingske}), and left (\textit{Information}) and right-wing political observation (\textit{Ekstrabladet}). This paper focuses on moderate newspapers during phase 1 of Covid-19, but all results have been confirmed for the full political spectrum as well as all national newspapers in Denmark (see Appendix \ref{appA} for details on methods and data).

To establish a baseline for novelty and resonance, we computed the pr. newspaper linear slope for resonance on novelty ($\mathbb{N} \times \mathbb{R}$) from December 01, 2019 to February 26, 2020 (the first case of Covid-19 in Denmark was registered on February 27, 2020). As can be observed from Figure \ref{fig:baseline}, the slopes are remarkably similar, indicating a medium to strong association between novelty and resonance ($M=0.56,~SD=0.06$) before the national outbreak of Covid-19. In the normal state of affairs, novelty and resonance therefore seems to be coupled such that novel news items tend to resonate more than overused and repetitive items and vice versa. This general news dynamic confirms the intuition that news media, all things being equal, maintain their relevance by propagating news.

The normal state of affairs was however severely disrupted by the rapid spread of Covid-19 and following lockdown on March 13. For \textit{Politiken} in Figure \ref{fig:poladapt} it is apparent that between the first case (`Virus') and the `Lockdown' there is a non-linear decrease in novelty (upper panel) which is countered by an inverted resonance response (middle panel). This initial decoupling of novelty and resonance does, in other words, coincide with the catastrophic Covid-19 event. While slightly less pronounced, the same decoupling patterns can be observed in \textit{Berlingske}, see Figure \ref{fig:beradapt}. In fact, decoupling of novelty and resonance can be observed in all national broadsheet newspapers irrespective of political observation and rare news phenomena where all news items are concerned with one event for an extended period of time ($>3$ weeks).  

An interesting contrast that may be politically motivated can be observed around the phase 1 `Opening' at April 15. While the center-left newspaper \textit{Politiken} was quite fast at re-establishing an almost normal state of affairs, the center-right \textit{Berlingske} showed a delayed linear novelty response for the remainder of the first phase. This is likely to reflect the level of support for the center-left government's decision making regarding lockdown and opening. The center-right newspaper seems to retain the decoupling for longer as they continue to focus on the economic and societal consequences of the government's Covid-19 policies.

The $\mathbb{N} \times \mathbb{R}$ computed on event-defined windows, e.g., `$Lockdown \rightarrow Opening$', can be used for simple change detection technique, see Figures \ref{fig:poladapt} and \ref{fig:beradapt} lower panels. As phase 1 unfolds, the decoupling is observed as loss of association between novelty and resonance. In the `$Virus \rightarrow Lockdown$' window (middle plot in lower panel), the slope approaches zero (for \textit{Politiken} the slope is actually negative), when the decoupling is most pronounced. The partial return to normal can be observed in the (dis-)similarity between `$ \rightarrow Wuhan$' and `$Opening \rightarrow $' (\textit{Politiken}: $0.65 \rightarrow 0.6$, and \textit{Berlingske}: $0.48 \rightarrow 0.36$).

\section*{Concluding Remarks}
In conclusion we propose the empirically-derived \textit{News Information Decoupling} (NID) principle, which states that in response to unexpected and dangerous temporally extended events, the ordinary information dynamics of news media are (initially) decoupled such that the content novelty decreases as media focus monotonically on the catastrophic event, but the resonant property of said content increases as its continued relevance propagate throughout the news information system. As such, NID behavior of the news information flow is a signature of catastrophes. At the level of methodology, this paper has illustrated how NID behavior of legacy media can be identified using non-linear adaptive filtering and windowed linear fits for resonance and novelty, and function provide input for change point detection algorithms. 

There are multiple research paths that could contribute further to our understanding of NID. First, the predictive value of NID should be validated in terms of crisis management; second evaluation of NID's event scope is necessary, for instance, does NID only apply to a small set of negative events or does it generalize to a substantially larger set of significant events (e.g., moon landing, fall of the Berlin wall); third, comparisons of information micro-behavior as a function of newspaper values, for instance, left vs. right-wing newspapers or tabloid vs. broadsheet newspapers; and finally, multilingual validation of NID.

\section*{Appendix}
\appendix

\section{Methods}\label{appA}

\subsection*{Data and Normalization}
The data set consists of all linguistic content (title and body text) from front pages of four Danish national newspapers \textit{Politiken}, \textit{Berlingske}, \textit{Information} and \textit{Ekstrabladet}. The newspapers were sampled during December 1, 2019 to June 1 2020. Content not produced by the newspaper, e.g., advertisements, was excluded from the sample. In order to normalize linguistic content, numerals and highly frequent function words were removed, and the remaining data were lemmatized and casefolded. Subsequently, the data were represented as a bag-of-words (BoW) model using latent Dirichlet allocation in order to generate a dense low-rank representation of each article. Note that with a few modifications to equations \eqref{eq:4} and \eqref{eq:5}, the approach works for any probabilistic or geometric vector-representation of documents. Novelty and resonance were estimated for in windows of one week ($w = 7)$.

\subsection*{Novelty and Resonance}
Two related information signals were extracted from the temporally sorted BoW model: \textit{Novelty} as an article $s^{(j)}$'s reliable difference from past articles $s^{(j-1)}, s^{(j-2)} , \dots ,s^{(j-w)}$ in window $w$:

\begin{equation}
\mathbb{N}_w (j) = \frac{1}{w} \sum_{d=1}^{w}  JSD (s^{(j)} \mid s^{(j - d)})\label{eq:1}
\end{equation}

and \textit{resonance} as the degree to which future articles $s^{(j+1)}, s^{(j+2)}, \dots , s^{(j+w)}$ conforms to article $s^{(j)}$'s novelty:

\begin{equation}
\mathbb{R}_w (j) = \mathbb{N}_w (j) - \mathbb{T}_w (j)\label{eq:2}
\end{equation}

where $\mathbb{T}$ is the \textit{transience} of $s^{(j)}$:

\begin{equation}
\mathbb{T}_w (j) = \frac{1}{w} \sum_{d=1}^{w}  JSD (s^{(j)} \mid s^{(j + d)})\label{eq:3}
\end{equation}

The novelty-resonance model was originally proposed in \cite{barron_individuals_2018}, but here we propose a symmetrized and smooth version by using the Jensen–Shannon divergence ($JSD$):

\begin{equation}
JSD (s^{(j)} \mid s^{(k)}) =  \frac{1}{2} D (s^{(j)} \mid M) + \frac{1}{2} D (s^{(k)} \mid M)\label{eq:4}
\end{equation}

with $M = \frac{1}{2} (s^{(j)} + s^{(k)})$ and $D$ is the Kullback-Leibler divergence:

\begin{equation}
D (s^{(j)} \mid s^{(k)}) = \sum_{i = 1}^{K} s_i^{(j)} \times \log_2 \frac{s_i^{(j)}}{s_i^{(k)}}\label{eq:5}
\end{equation}

\subsection*{Nonlinear Adaptive Filtering}
To model global trends in the novelty and resonance signals, we apply a nonlinear adaptive multi-scale decomposition algorithm \cite{gao_facilitating_2011}. First, the signal is partitioned into overlapping segments of length $w = 2n+1$, where neighboring segments overlap by $n+1$ points. In each segment, the signal is fitted with the best polynomial of order $M$, obtained by using the standard least-squares regression; the fitted polynomials in overlapped regions are then combined to yield a single global smooth trend. Denoting the fitted polynomials for the $i-th$ and $(i+1)-th$ segments by $y^{i} (l_1)$ and $y^{(i+1)} (l_2)$, respectively, where $l_1, l_2 = 1,\cdots,2n+1$, we define the fitting for the overlapped region as

\begin{equation}
y^{(c)} (l) = w_1 y^{(i)} (l+n) + w_2 y^{(i+1)} (l),~~l=1,2,\cdots,n+1\label{eq:6}
\end{equation}

\noindent where $w_1 = \big (1-\frac{l-1}{n} \big )$ and $w_2=\frac{l-1}{n}$ can be written as $(1-d_j/n)$ for $j=1,2$, and where $d_j$ denotes the distances between the point and the centers of $y^{(i)}$ and $y^{(i+1)}$, respectively. Note that the weights decrease linearly with the distance between the point and the center of the segment. Such a weighting is used to ensure symmetry and effectively eliminate any jumps or discontinuities around the boundaries of neighboring segments. As a result, the global trend is smooth at the non-boundary points, and has the right and left derivatives at the boundary \cite{riley_tutorial_2012}. 

The global trend thus determined can be used to maximally suppress the effect of complex nonlinear trends on the scaling analysis. The parameters of each local fit is determined by maximizing the goodness of fit in each segment. The different polynomials in overlapped part of each segment are combined using Equation \ref{eq:6} so that the global fit will be the best (smoothest) fit of the overall time series. Note that, even if $M = 1$ is selected, i.e., the local fits are linear, the global trend signal will still be nonlinear.

Finally, in order to describe the information states before and after an events (e.g., Lockdown, Opening), we fit resonance on novelty to estimate the $\mathbb{N}\times\mathbb{R}$ slope $\beta_1$ in the specific time windows:

\begin{equation}
\mathbb{R}_i = \beta_0 + \beta_1 \mathbb{N}_i + \epsilon_i, ~~ i = 1, \dots, n.\label{eq:7}
\end{equation}

\section{Online Resources}

All data are proprietary and have been collected through Infomedia's API: \url{https://infomedia.dk/}. For inquiries regarding models and derived data, please contact \url{kln@cas.au.dk}. The source code is available on Github: \url{https://bit.ly/3beahFd}. MOre details on NID detection can be found at NeiC's NDHL website: \url{https://bit.ly/3bfeW9C}.

\begin{acknowledgments}
This paper has been supported the "HOPE - How Democracies Cope with COVID-19"-project funded by The Carlsberg Foundation with grant CF20-0044 and NeiC's Nordic Digital Humanities Laboratory project. The authors would like to thank Infomedia for access to proprietary data.
\end{acknowledgments}

\bibliography{main}

\begin{thebibliography}{17}
\expandafter\ifx\csname natexlab\endcsname\relax\def\natexlab#1{#1}\fi
\providecommand{\url}[1]{\texttt{#1}}
\providecommand{\href}[2]{#2}
\providecommand{\path}[1]{#1}
\providecommand{\DOIprefix}{doi:}
\providecommand{\ArXivprefix}{arXiv:}
\providecommand{\URLprefix}{URL: }
\providecommand{\Pubmedprefix}{pmid:}
\providecommand{\doi}[1]{\href{http://dx.doi.org/#1}{\path{#1}}}
\providecommand{\Pubmed}[1]{\href{pmid:#1}{\path{#1}}}
\providecommand{\bibinfo}[2]{#2}
\ifx\xfnm\relax \def\xfnm[#1]{\unskip,\space#1}\fi
\bibitem[{Guldi(2019)}]{guldi_measures_2019}
\bibinfo{author}{J.~Guldi},
\newblock \bibinfo{title}{The {Measures} of {Modernity}: {The} {New}
  {Quantitative} {Metrics} of {Historical} {Change} {Over} {Time} and {Their}
  {Critical} {Interpretation}},
\newblock \bibinfo{journal}{International Journal for History, Culture and
  Modernity} \bibinfo{volume}{7} (\bibinfo{year}{2019})
  \bibinfo{pages}{899--939}. \URLprefix
  \url{https://brill.com/view/journals/hcm/7/1/article-p899_42.xml}.
  \DOIprefix\doi{10.18352/hcm.589}.
\bibitem[{van Eijnatten and Ros(2019)}]{van_eijnatten_eurocentric_2019}
\bibinfo{author}{J.~van Eijnatten}, \bibinfo{author}{R.~Ros},
\newblock \bibinfo{title}{The {Eurocentric} {Fallacy}. {A}
  {Digital}-{Historical} {Approach} to the {Concepts} of ‘{Modernity}’,
  ‘{Civilization}’ and ‘{Europe}’ (1840–1990)},
\newblock \bibinfo{journal}{International Journal for History, Culture and
  Modernity} \bibinfo{volume}{7} (\bibinfo{year}{2019})
  \bibinfo{pages}{686--736}. \URLprefix
  \url{https://brill.com/view/journals/hcm/7/1/article-p686_33.xml}.
  \DOIprefix\doi{10.18352/hcm.580}.
\bibitem[{Daems et~al.(2019)Daems, D’haeninck, Hengchen, Zere, and
  Verbruggen}]{daems_workers_2019}
\bibinfo{author}{J.~Daems}, \bibinfo{author}{T.~D’haeninck},
  \bibinfo{author}{S.~Hengchen}, \bibinfo{author}{T.~Zere},
  \bibinfo{author}{C.~Verbruggen},
\newblock \bibinfo{title}{‘{Workers} of the {World}’? {A} {Digital}
  {Approach} to {Classify} the {International} {Scope} of {Belgian} {Socialist}
  {Newspapers}, 1885–1940},
\newblock \bibinfo{journal}{Journal of European Periodical Studies}
  \bibinfo{volume}{4} (\bibinfo{year}{2019}) \bibinfo{pages}{99--114}.
  \URLprefix \url{https://ojs.ugent.be/jeps/article/view/10187}.
  \DOIprefix\doi{10.21825/jeps.v4i1.10187}.
\bibitem[{Kestemont et~al.(2014)Kestemont, Karsdorp, and
  Düring}]{kestemont_mining_2014}
\bibinfo{author}{M.~Kestemont}, \bibinfo{author}{F.~Karsdorp},
  \bibinfo{author}{M.~Düring},
\newblock \bibinfo{title}{Mining the {Twentieth} {Century}’s {History} from
  the {Time} {Magazine} {Corpus}},
\newblock in: \bibinfo{booktitle}{Proceedings of the 8th {Workshop} on
  {Language} {Technology} for {Cultural} {Heritage}, {Social} {Sciences}, and
  {Humanities} ({LaTeCH})}, \bibinfo{publisher}{Association for Computational
  Linguistics}, \bibinfo{address}{Gothenburg, Sweden}, \bibinfo{year}{2014},
  pp. \bibinfo{pages}{62--70}. \URLprefix
  \url{http://aclweb.org/anthology/W14-0609}.
  \DOIprefix\doi{10.3115/v1/W14-0609}.
\bibitem[{Bos et~al.(2016)Bos, Wijfjes, Piscaer, and
  Voerman}]{bos_quantifying_2016}
\bibinfo{author}{P.~Bos}, \bibinfo{author}{H.~Wijfjes},
  \bibinfo{author}{M.~Piscaer}, \bibinfo{author}{G.~Voerman},
\newblock \bibinfo{title}{Quantifying “{Pillarization}”: {Extracting}
  {Political} {History} from {Large} {Databases} of {Digitized} {Media}
  {Collections}},
\newblock \bibinfo{journal}{Proceedings of the 3rd HistoInformatics Workshop}
  (\bibinfo{year}{2016}) \bibinfo{pages}{10}.
\bibitem[{Newman and Block(2006)}]{newman_probabilistic_2006}
\bibinfo{author}{D.~J. Newman}, \bibinfo{author}{S.~Block},
\newblock \bibinfo{title}{Probabilistic topic decomposition of an
  eighteenth-century {American} newspaper},
\newblock \bibinfo{journal}{Journal of the American Society for Information
  Science and Technology} \bibinfo{volume}{57} (\bibinfo{year}{2006})
  \bibinfo{pages}{753--767}. \URLprefix
  \url{http://doi.wiley.com/10.1002/asi.20342}.
  \DOIprefix\doi{10.1002/asi.20342}.
\bibitem[{Wevers(2019)}]{wevers_using_2019}
\bibinfo{author}{M.~Wevers},
\newblock \bibinfo{title}{Using {Word} {Embeddings} to {Examine} {Gender}
  {Bias} in {Dutch} {Newspapers}, 1950-1990},
\newblock in: \bibinfo{booktitle}{Proceedings of the 1st {International}
  {Workshop} on {Computational} {Approaches} to {Historical} {Language}
  {Change}}, \bibinfo{publisher}{Association for Computational Linguistics},
  \bibinfo{address}{Florence, Italy}, \bibinfo{year}{2019}, pp.
  \bibinfo{pages}{92--97}. \URLprefix
  \url{https://www.aclweb.org/anthology/W19-4712}.
  \DOIprefix\doi{10.18653/v1/W19-4712}.
\bibitem[{Wevers et~al.(2020)Wevers, Gao, and Nielbo}]{wevers_tracking_2020}
\bibinfo{author}{M.~Wevers}, \bibinfo{author}{J.~Gao}, \bibinfo{author}{K.~L.
  Nielbo},
\newblock \bibinfo{title}{Tracking the {Consumption} {Junction}: {Temporal}
  {Dependencies} between {Articles} and {Advertisements} in {Dutch}
  {Newspapers}},
\newblock \bibinfo{journal}{Digital Humanities Quarterly} \bibinfo{volume}{014}
  (\bibinfo{year}{2020}).
\bibitem[{Gao et~al.(2012)Gao, Hu, Mao, and Perc}]{gao_culturomics_2012}
\bibinfo{author}{J.~Gao}, \bibinfo{author}{J.~Hu}, \bibinfo{author}{X.~Mao},
  \bibinfo{author}{M.~Perc},
\newblock \bibinfo{title}{Culturomics meets random fractal theory: insights
  into long-range correlations of social and natural phenomena over the past
  two centuries},
\newblock \bibinfo{journal}{Journal of The Royal Society Interface}
  \bibinfo{volume}{9} (\bibinfo{year}{2012}) \bibinfo{pages}{1956--1964}.
\bibitem[{Barron et~al.(2018)Barron, Huang, Spang, and
  DeDeo}]{barron_individuals_2018}
\bibinfo{author}{A.~T.~J. Barron}, \bibinfo{author}{J.~Huang},
  \bibinfo{author}{R.~L. Spang}, \bibinfo{author}{S.~DeDeo},
\newblock \bibinfo{title}{Individuals, institutions, and innovation in the
  debates of the {French} {Revolution}},
\newblock \bibinfo{journal}{Proceedings of the National Academy of Sciences}
  \bibinfo{volume}{115} (\bibinfo{year}{2018}) \bibinfo{pages}{4607--4612}.
  \URLprefix \url{http://www.pnas.org/lookup/doi/10.1073/pnas.1717729115}.
  \DOIprefix\doi{10.1073/pnas.1717729115}.
\bibitem[{Murdock et~al.(2015)Murdock, Allen, and
  DeDeo}]{murdock_exploration_2015}
\bibinfo{author}{J.~Murdock}, \bibinfo{author}{C.~Allen},
  \bibinfo{author}{S.~DeDeo},
\newblock \bibinfo{title}{Exploration and {Exploitation} of {Victorian}
  {Science} in {Darwin}'s {Reading} {Notebooks}},
\newblock \bibinfo{journal}{arXiv preprint arXiv:1509.07175}
  (\bibinfo{year}{2015}). \URLprefix \url{http://arxiv.org/abs/1509.07175}.
\bibitem[{Nielbo et~al.(2019{\natexlab{a}})Nielbo, Perner, Larsen, Nielsen, and
  Laursen}]{nielbo_automated_2019}
\bibinfo{author}{K.~L. Nielbo}, \bibinfo{author}{M.~L. Perner},
  \bibinfo{author}{C.~P. Larsen}, \bibinfo{author}{J.~Nielsen},
  \bibinfo{author}{D.~Laursen},
\newblock \bibinfo{title}{Automated {Compositional} {Change} {Detection} in
  {Saxo} {Grammaticus}' {Gesta} {Danorum}},
\newblock in: \bibinfo{booktitle}{{DHN}}, \bibinfo{year}{2019}{\natexlab{a}},
  pp. \bibinfo{pages}{320--332}.
\bibitem[{Nielbo et~al.(2019{\natexlab{b}})Nielbo, Baunvig, Liu, and
  Gao}]{nielbo_curious_2019}
\bibinfo{author}{K.~L. Nielbo}, \bibinfo{author}{K.~F. Baunvig},
  \bibinfo{author}{B.~Liu}, \bibinfo{author}{J.~Gao},
\newblock \bibinfo{title}{A curious case of entropic decay: {Persistent}
  complexity in textual cultural heritage},
\newblock \bibinfo{journal}{Digital Scholarship in the Humanities}
  \bibinfo{volume}{34} (\bibinfo{year}{2019}{\natexlab{b}}). \URLprefix
  \url{https://doi.org/10.1093/llc/fqy054}. \DOIprefix\doi{10.1093/llc/fqy054}.
\bibitem[{Nielbo et~al.(2021)Nielbo, Vahlstrup, and
  Bechmann}]{nielbo_trend_2021}
\bibinfo{author}{K.~Nielbo}, \bibinfo{author}{P.~Vahlstrup},
  \bibinfo{author}{A.~Bechmann},
\newblock \bibinfo{title}{Trend {Reservoir} {Detection}: {Minimal}
  {Persistence} and {Resonant} {Behavior} of {Trends} in {Social} {Media}},
\newblock \bibinfo{journal}{Proceedings of Computational Humanities Research}
  \bibinfo{volume}{1} (\bibinfo{year}{2021}).
\bibitem[{Vrangbæk and Nielbo(2021)}]{vrangbaek_composition_2021}
\bibinfo{author}{E.~Vrangbæk}, \bibinfo{author}{K.~Nielbo},
\newblock \bibinfo{title}{Composition and {Change} in {De} {Civitate} {Dei}:
  {A} {Case} {Study} of {Computationally} {Assisted} {Methods}},
\newblock \bibinfo{journal}{Studia Patristica}  (\bibinfo{year}{2021}).
\bibitem[{Gao et~al.(2011)Gao, Hu, and Tung}]{gao_facilitating_2011}
\bibinfo{author}{J.~Gao}, \bibinfo{author}{J.~Hu}, \bibinfo{author}{W.-w.
  Tung},
\newblock \bibinfo{title}{Facilitating {Joint} {Chaos} and {Fractal} {Analysis}
  of {Biosignals} through {Nonlinear} {Adaptive} {Filtering}},
\newblock \bibinfo{journal}{PLoS ONE} \bibinfo{volume}{6}
  (\bibinfo{year}{2011}) \bibinfo{pages}{e24331}. \URLprefix
  \url{http://dx.plos.org/10.1371/journal.pone.0024331}.
  \DOIprefix\doi{10.1371/journal.pone.0024331}.
\bibitem[{Riley et~al.(2012)Riley, Bonnette, Kuznetsov, Wallot, and
  Gao}]{riley_tutorial_2012}
\bibinfo{author}{M.~A. Riley}, \bibinfo{author}{S.~Bonnette},
  \bibinfo{author}{N.~Kuznetsov}, \bibinfo{author}{S.~Wallot},
  \bibinfo{author}{J.~Gao},
\newblock \bibinfo{title}{A tutorial introduction to adaptive fractal
  analysis},
\newblock \bibinfo{journal}{Frontiers in Physiology} \bibinfo{volume}{3}
  (\bibinfo{year}{2012}). \URLprefix
  \url{http://journal.frontiersin.org/article/10.3389/fphys.2012.00371/abstract}.
  \DOIprefix\doi{10.3389/fphys.2012.00371}.

\end{thebibliography}
\end{document}